\documentclass[twocolumn,aps,prl,amsfonts,showpacs,reprint,floatfix]{revtex4-1}
\usepackage{amsmath}
\usepackage{amsfonts}
\usepackage{graphicx}
\usepackage{color}
\begin{document}

\title{An interface phase transition induced by a driven line in 2D}

\author{Tridib Sadhu}
\email{tsadhu@gmail.com}

\author{Zvi Shapira}
\email{zvi.shapira@weizmann.ac.il}

\author{David Mukamel}
\email{david.mukamel@weizmann.ac.il}

\affiliation{Department of physics of complex systems, Weizmann institute of science,
Rehovot-76100, Israel.}
\begin{abstract}
The effect of a localized drive on the steady state of an interface separating two
phases in coexistence is studied. This is done using a
spin conserving kinetic Ising model on a two dimensional lattice with cylindrical
boundary conditions, where a drive is applied along a
single ring on which the interface separating the two phases is centered. The drive
is found to induce an interface spontaneous symmetry breaking whereby the magnetization
of the driven ring becomes non-zero. The width of the interface becomes finite and its
fluctuations around the driven ring are non-symmetric.
The dynamical origin of these properties is analyzed in an adiabatic limit which allows the evaluation
of the large deviation function of the driven-ring magnetization.
\end{abstract}

\maketitle
The effect of local drive on the properties of an interface
separating two coexisting phases has recently been explored as a
simple example of systems driven away from equilibrium. Much of the
attention is due to the surprising experimental results
on colloidal gas-liquid interface subjected to a shear flow
parallel to the interface \cite{DERKS}. It was found that the shear drive applied
away from the interface, strongly suppresses the fluctuations of the
interface, making it smoother. This long-distance effect of the drive
is due to long-range correlations that characterize
driven systems \cite{SMM,KLS1,*KLS2,SPOHN,ZHANG,GARRIDO,RUBI,ZIA}.
An interesting theoretical approach
for studying this phenomenon has been introduced by Smith \textit{et
al.} who
considered a two dimensional version of the system, and modeled it by
an Ising lattice-gas below its transition temperature
\cite{SMITH}.
Using spin conserving Kawasaki dynamics and applying shear
flow at the boundaries parallel to the interface, it was observed that the interface
indeed becomes narrower although its width still increases with the
length of the interface. In closely related works, the effect induced by a current carrying line
on a neighboring non-driven one has also been analyzed \cite{DICKMAN,KOLOMEISKY,HUCHT,HILHORST}.

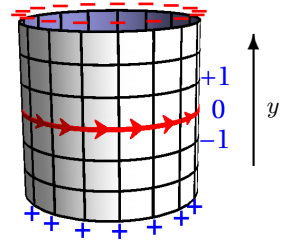
\begin{figure}[t]
\begin{center}
\includegraphics[width=3.0cm]{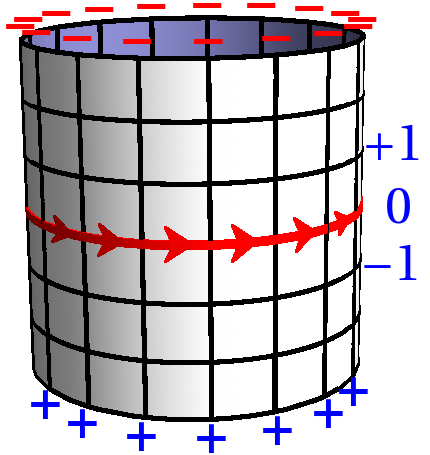}
\caption{(color online) Square lattice with cylindrical boundary
condition, with the drive on the central ring and
the boundary conditions indicated.\label{fig:Ising-spin-on}}
\end{center}
\begin{picture}(0,0)
\thicklines
\put(50,90){\vector(0,1){50}}
\put(55,110){$y$}
\end{picture}
\end{figure}

In this Letter we consider a drive localized \textit{along} an interface
which separates two coexisting phases, and study the resulting interface properties. This is done using a two
dimensional Ising model on a square lattice with cylindrical boundary
conditions (Fig. \ref{fig:Ising-spin-on}), that evolves under spin conserving dynamics. The drive acts along the
ring around which the interface is centered.
We find that the drive induces an interface phase transition which
involves spontaneous symmetry breaking, resulting in a non-zero
magnetization of the driven ring.
In this transition, the macroscopic $2D$ steady state remains
unchanged,
however spontaneous symmetry breaking takes place involving the steady state
of a $1D$ stripe centered on the driven ring.
This is in sharp contrast with an equilibrium setup of an interface
subjected to a localizing potential along a ring, where the ring magnetization
vanishes at all temperatures, and no interface spontaneous symmetry breaking
takes place. Moreover, we find that the drive suppresses
the fluctuations of the interface, leading to an interface with a
finite width which does not scale with the system size. Also, due to
the broken symmetry on the driven ring, the interface fluctuations
are highly asymmetric. The interface
fluctuates more strongly into the bulk phase whose magnetization is
opposite to that of the driven ring. These results are first
demonstrated by numerical simulations. The model is then analyzed
in a special limit which allows analytical computation of the large
deviation function (LDF) \cite{TOUCHETTE} of the magnetization of the driven ring,
demonstrating the existence of the spontaneous symmetry breaking.

To proceed, we consider Ising spins $\boldsymbol{\sigma}\equiv\left\{
\sigma_{\mathbf{r}}\right\} $ on sites $\mathbf{r}=(x,y)$ of an
$L\times(2M+1)$ square lattice, with periodic boundary condition in the x-direction
while the two open boundaries, $y=\pm M$, are coupled to rows from
above ($y=M+1$) and below ($y=-M-1$), respectively, with fixed spins:
$\sigma_{x,\pm M\pm1}=\mp 1$ (Fig. \ref{fig:Ising-spin-on}). 
The model has nearest neighbor ferromagnetic interactions
and a drive is introduced by a force field $\mathbf{E}\equiv(E,0)$ applied
on the $y=0$ ring. The field favors the positive spins to move counter
clockwise along the ring, and as a result drives the system out of equilibrium.
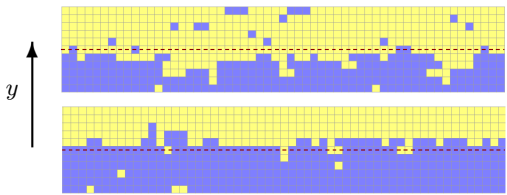
\begin{figure}
\begin{center}
\includegraphics[width=6.0cm]{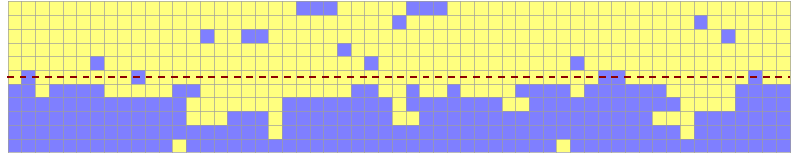}\\
\vspace{1.5mm}
\includegraphics[width=6.0cm]{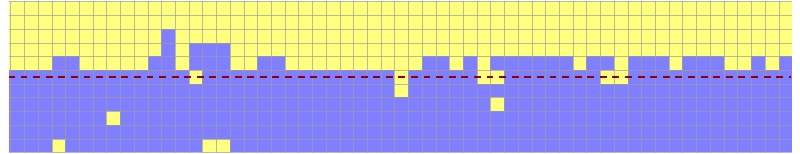}
\caption{(color online) Typical spin configurations zoomed around
the driven ring (denoted by a dashed line) for $L=2M=100$ at
$T=0.85T_{c}$. Blue (dark) and yellow (light) squares
indicate $+$ and $-$ spins, respectively.
\label{fig:An-example-of}}
\end{center}
\begin{picture}(0,0)
\thicklines
\put(-95,90){\vector(0,1){40}}
\put(-105,110){$y$}
\end{picture}
\end{figure}

There is more than one way to incorporate the drive in the dynamics, and unlike
a dynamics satisfying detailed balance, the steady state
depends on the precise choice of the rates \cite{KWAK}. We choose a modified Metropolis
algorithm \cite{KLS1} where in every step a pair of nearest neighbor
sites $\mathbf{r}$ and $\mathbf{r'}$ are chosen at random and their
spins are exchanged with probability $\rm{min}\{1,\exp(-\beta\Delta\mathcal{H})\}$,
where $\beta$ is an inverse temperature and $\Delta\mathcal{H}$
is the energy difference between the final and initial configurations.
Thus, for exchanging $\sigma_{\mathbf{r}}$ and
$\sigma_{\mathbf{r^{\prime}}}$,
\begin{equation}
\Delta\mathcal{H}=\begin{cases}\Delta
H-(\sigma_{\mathbf{r}}-\sigma_{\mathbf{r^{\prime}}})(\mathbf{r^{\prime}}-\mathbf{r})\cdot\mathbf{E}
& \text{if \ensuremath{\mathbf{r}},\ensuremath{\mathbf{r'\in}}
\ensuremath{0}th ring,}\\
\Delta H & \text{elsewhere,}
\end{cases}\nonumber\label{eq:rates}
\end{equation}
where $\Delta H$ is calculated using the Ising Hamiltonian
$H=-J\sum_{\langle
\mathbf{r},\mathbf{r'}\rangle}\sigma_{\mathbf{r}}\sigma_{\mathbf{r^{\prime}}}$,
with $J>0$. One Monte Carlo time step is constituted of
$L(4M+1)$ such updates. In all the numerical results presented in this
Letter we use a large driving field $E\ge10 J$.

In absence of a driving field, the model is
in equilibrium. At sub-critical temperatures ($T<T_{c}\simeq2.2692J/k_{B})$,
the equilibrium state is composed of two oppositely magnetized phases, separated by
an interface. For an initial configuration with zero overall
magnetization, the magnetization profile in the $y$ direction, $m_y \equiv 1/L\sum_x\sigma_{x,y}$,
is antisymmetric with respect to $y=0$. The interface fluctuates
symmetrically around the driven ring, leading to zero
magnetization on the ring, $ m_0=0$. In the large $L,M$ limit with fixed aspect
ratio $L/M$, the width of the interface scales as $\sqrt L$
\cite{REED}.

Introducing a drive does not modify the overall macroscopic structure of the steady state.
As is naively expected, the steady state is still composed of two oppositely magnetized
phases separated by a fluctuating interface around $y=0$. However, numerical studies of
the model reveal some profound changes in the structure of the
interface itself. In particular we find
that (a) in the thermodynamic limit the magnetization of the
driven line, $ m_0 $,
is non-zero, taking one of two oppositely directed values. It thus breaks the
$ \sigma_{\mathbf{r}}\rightarrow -\sigma_{-\mathbf{r}}$
symmetry of the model. (b) The interface is localized around the driven line and
its width stays finite in the thermodynamic limit, and (c) the fluctuations of the interface
into the two bulk phases are highly asymmetric, with more pronounced fluctuations into the phase
whose magnetization is oppositely directed to that of the driven line.

\begin{figure}
\begin{center}
\includegraphics[width=5.5cm]{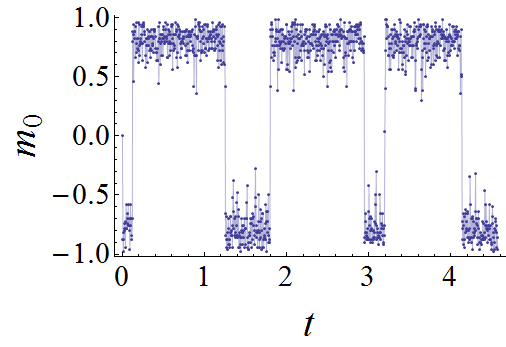}
\begin{picture}
(0,0){\put(1.0,18.0){\small $\times10^{7}$}}
\end{picture}
\caption{Typical time evolution of $m_{0}$ for $L=2M=100$ at
$T=0.6T_{c}$. Time is measured in Monte Carlo steps. \label{fig:Time-evolution-of}}
\end{center}
\end{figure}
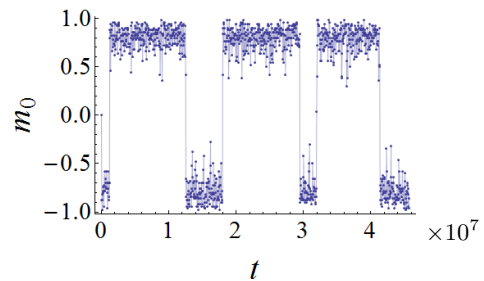

In Fig. \ref{fig:An-example-of} we present two typical microscopic
configurations of the model. It is clearly seen that the driven line
is predominantly occupied by either positive or negative spins representing its two possible
ordered states.
As the system evolves, the magnetization
$m_0$ fluctuates around one of the non-zero values for a long
time. It then switches to the oppositely magnetized state
over a much shorter time scale, as shown in Fig. \ref{fig:Time-evolution-of}.

\begin{figure}
\begin{center}
\includegraphics[width=5.3cm]{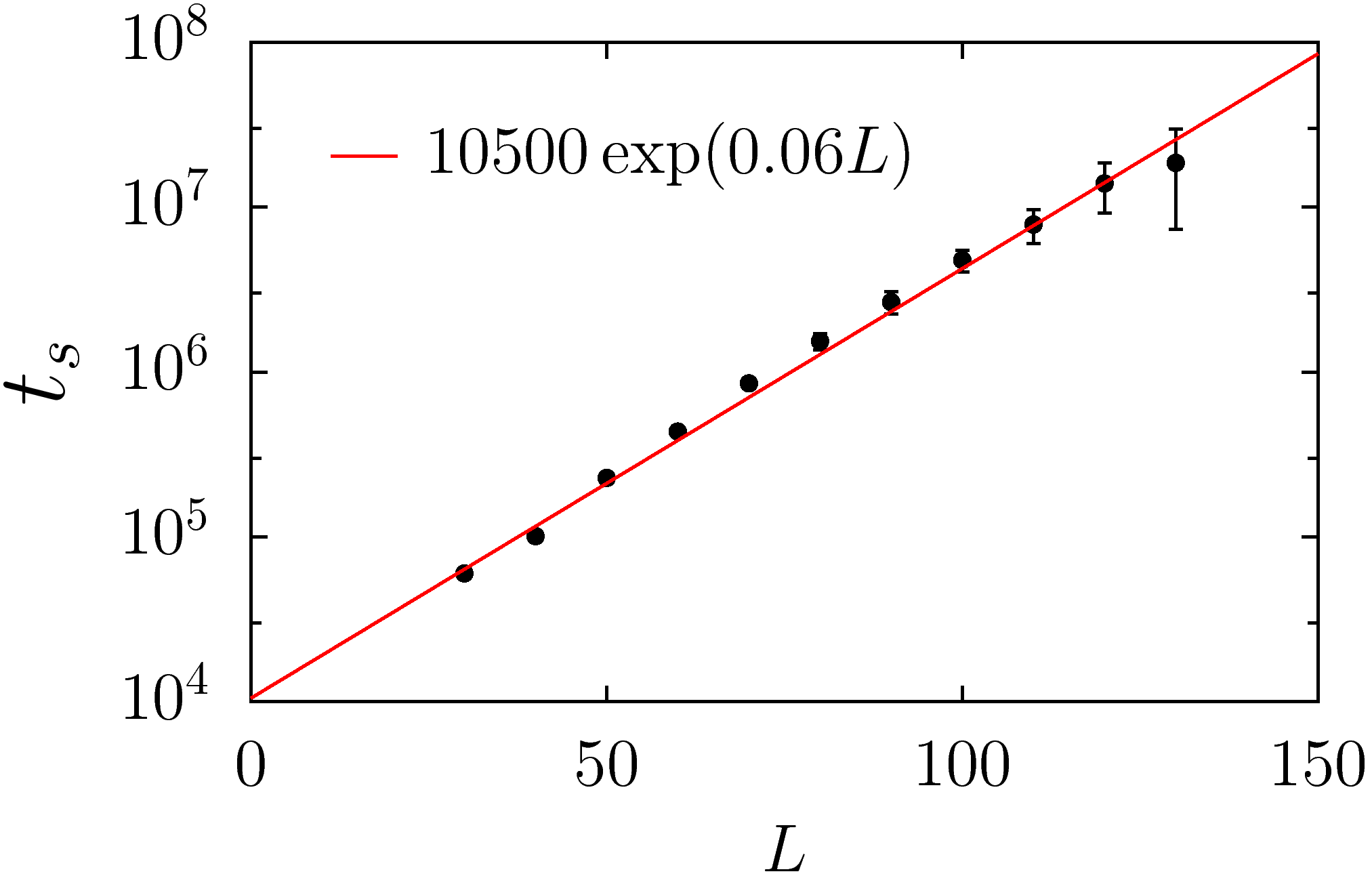}
\caption{Average time between consecutive switches $t_{s}(L)$ for
$2M=L$, and $T=0.6T_{c}$ \label{fig:Exponential-growth-of}}
\end{center}
\end{figure}

The numerical result for the average time between two successive such
switches, $t_s$, are shown in
Fig. \ref{fig:Exponential-growth-of}. The data suggest that $t_s$
grows exponentially with $L$,  with $t_s \sim \exp(0.06L)$.
The data for each $L$ is averaged over $n$ number of switches that are
observed in available computation time ($n$ varies from around $12000$ to
$10$ as $L$ changes from $30$ to $130$); $n$ decreases with $L$,
yielding increasing error bars of order $1/\sqrt{n}$ with $L$.
Although the range of the system size studied is
insufficient for a conclusive evidence of an exponential growth, this form is justified by
the theoretical results presented below. The exponential growth implies that in the
thermodynamic limit, the two non-zero values of $m_0$ correspond to two thermodynamically
stable phases.
\begin{figure}
\begin{center}
\includegraphics[width=6.0cm]{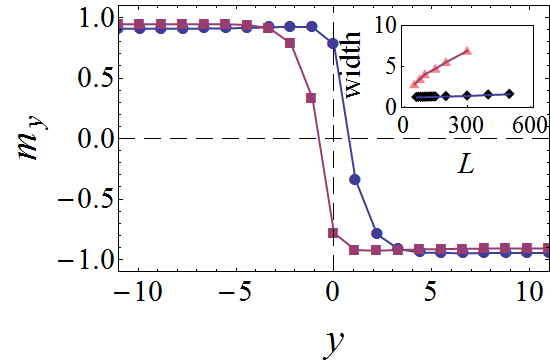}
\caption{(color online) The average magnetization profile $m_{y}$ corresponding to the two phases, close to
$y=0$. The asymmetry around $y=0$ is clearly seen. The profiles
are generated on a $100\times101$ lattice at $T=0.85T_{c}$,
averaging over $10^{5}$ configurations at regular
intervals of $1000$ Monte Carlo steps. The figure in the inset shows
the growth of the width of the interface with increasing length
$L$ for zero drive (top curve) and for driving strength $E=10 J$
(lower curve). \label{fig:The-average-magnetization}}
\end{center}
\end{figure}

The width of the interface is evaluated by averaging $\vert y\vert$
weighted by the derivative $dm_{y}/dy$ that peaks at the interface
position. The result is shown in the inset of Fig.
\ref{fig:The-average-magnetization} for both driven and non-driven
case. A comparison of the two cases clearly indicates that
the interface fluctuations are drastically reduced in the presence of drive.
As will be shown by the theoretical analysis presented below, the width of the interface
remains finite at large $L$. Such
smoothening of the interface has also been observed in presence of global drive parallel to the
interface \cite{LEUNG1,*LEUNG2}. The interesting difference here is
that the interfacial fluctuations are
asymmetric, resulting in an asymmetric magnetization profile around the driven line (see
Fig. \ref{fig:The-average-magnetization}).

\begin{figure}
\includegraphics[width=6.0cm]{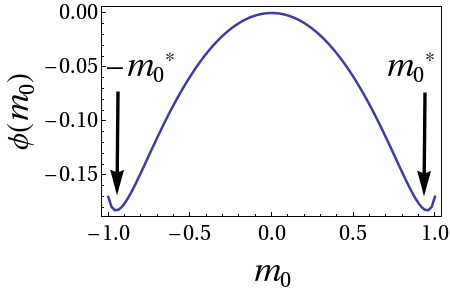}
\caption{The large deviation function $\phi(m_{0})$ calculated using
Eq. \eqref{eq:potential} and \eqref{eq:rate}. \label{fig:The-potential-function}}
\end{figure}

In order to make an analytical analysis of the model feasible, we generalize
the model by introducing a parameter $\gamma$ that controls the dynamical
rate of the processes involving spin exchange between the $y=0$ ring
and the neighboring rings $y=\pm 1$. For these processes the rate
becomes $\rm{min}\{\gamma,\gamma\exp\left(-\beta
\Delta\mathcal{H}\right)\}$, with $\gamma>0$. The other rates remain
unchanged. This does not modify the steady state
of the equilibrium case ($E=0$) but it helps analyzing the non-equilibrium steady state.
We now consider the steady state in the following special limit: (a) slow exchange rates
$(\gamma \ll L^{-3})$ between the driven and the neighboring rings, (b) an infinite driving
field $\left(E\rightarrow\infty\right)$, and (c) low temperature $(\exp(-\beta J)\ll1)$.
We show below that in this limit the stationary probability
distribution $P\left( m_{0} \right)$ of the magnetization $m_{0}$ of
the driven line has the form  $P\left( m_{0} \right)\propto
\exp(-L \phi(m_{0}))$. The large deviation function,
$\phi(m_{0})$, is then computed and shown to possess two
degenerate minima at non-vanishing values of the magnetization
$m_{0}=\pm m_{0}^{\star}$ (see Fig. \ref{fig:The-potential-function}), implying
a spontaneous symmetry breaking on the ring. In addition, the LDF yields an exponential flipping
time in $L$ between positive and negative magnetization for finite systems due to the finite barrier
between the two minima.

We proceed by noting that due to the slow exchange rate $\gamma$, there are no significant exchanges between the driven
line and its neighboring rings on a time scale $t \ll \Delta t= (\gamma
L)^{-1}$. On such time scale the
lattice may be considered as composed of
three subsystems: the driven line, and the upper $u$ ($y>0$)
and lower $\ell$ ($y<0$) sublattices. They evolve while
keeping their own specific magnetization $m_0$, $m_{u}$, and $m_{\ell}$
unchanged, reaching the steady state corresponding to fixed
subsystem magnetization. On a longer timescale, $t\gtrsim \Delta t$,
the magnetizations $m_0$, $m_{u}$, and $m_{\ell}$
evolve as spins are exchanged between the subsystems.

We now define a coarse-grained time variable $\tau=t/\Delta t$
such that the subsystem magnetization evolves with increasing
$\tau$, however at any given $\tau$ each subsystem is effectively
in the steady state corresponding to its magnetization.
This separation of slow and fast processes is analogous to the
adiabatic approximation in quantum mechanics \cite{ADIABATIC},
and has also been applied in related models \cite{BEIJEREN,COHEN}.

Let us characterize the steady states corresponding to fixed sub-system
magnetization $m_{0}$, $m_{u}$ and $m_{\ell}$. First consider the driven ring. In the
limit $E\rightarrow\infty$ the dynamics within this ring is
independent of the two other subsystems, and reduces to that of the
Totally Asymmetric Simple Exclusion Process (TASEP). In its steady state all spin
configurations with fixed magnetization $m_0$ are equally probable, leading to uniform
magnetization and zero spin-spin correlation along the driven line.
This steady state is reached in a time of $\mathcal{O}(L^{3/2})$
which, for $\gamma\ll L^{-3}$, is smaller
than the typical time of exchange processes between the driven ring
and its neighboring ones.
Then, the driven ring provides an effective boundary magnetic field $J m_0$
on the $u$ and $\ell$ subsystems. Thus the steady state of these two subsystems is the equilibrium
state of the Ising model subjected to a boundary field. The boundary field results a magnetization profile,
$m_y$, which for large $|y|$ approaches the bulk magnetization values $-m_B$ and $m_B$ for the $u$ and $\ell$
subsystems, respectively. The length scale of this approach is of the order of the spin-spin correlation length $\xi(T)$
of the $2D$ Ising model. Since this length is finite at all temperatures except at $T_c$ this demonstrates that
the width of the interface remains finite for large $L$. 

Let $P_{\tau}(m_{0})d\tau$ be the probability of the driven line
magnetization to have value $m_{0}$ between coarse-grained time $\tau$
and $\tau+d\tau$, while $m_{u}$ and $m_{\ell}$ have already reached
stationary values. The probability function evolves as spins are
exchanged between the subsystems. At each exchange process between the
driven ring and the bulk, $m_{0}$ changes by $\pm 2/L$. Let $p(m_{0})$ and $q(m_{0})$ be the
increasing and decreasing rates of $m_{0}$, respectively. Then the
dynamics of $m_{0}$ is that of a random walker with position dependent
forward and backward jump rates $p(m_{0})$ and $q(m_{0})$,
respectively, and with boundary condition $P_{\tau}(m_{0})=0$ for $\vert m_{0} \vert > 1$.

The stationary distribution of this motion is an equilibrium distribution function,
\begin{equation}
P(m_{0})=P(0)\exp\left[-L \phi\left(m_{0}\right)\right].
\end{equation}
The LDF $\phi(m_{0})$ is an even function of $m_0$ and it can be determined using the detailed
balance condition $p(m_{0})P(m_{0})=q(m_{0}+2/L)P(m_{0}+2/L)$, which for
$0<m_{0}\le 1$ yields
\begin{equation}
P\left(m_{0}=\frac{2n}{L}\right)=P\left( 0
\right)\prod_{k=1}^{n}\frac{p\left[\frac{2}{L}(k-1)\right]}{q\left[
 \frac{2}{L}k\right]},
\end{equation}
with $n=1,\dots, L/2$.
In the large $L$ limit this yields the LDF for
$m\ge 0$,
\begin{equation}
\phi(m_{0})=\frac{1}{2}\int_{0}^{m_{0}}dm\ln\left[\frac{p\left(
m \right)}{q\left( m \right)}
\right],\label{eq:potential}
\end{equation}
with $\phi(-m_{0})=\phi(m_{0})$.

The rates $p\left( m_{0} \right)$ and $q\left( m_{0} \right)$ are
determined as follows: consider a spin exchange process between the driven ring and its
two neighboring ones, in which the microscopic configuration changes
from $\boldsymbol{\sigma}$ to
$\boldsymbol{\sigma^{\prime}}$ and $m_{0}$ increases by $2/L$.  The rate of this
process is $\omega\left( \boldsymbol{\sigma} \rightarrow \boldsymbol{\sigma'}
\right)P\left(\boldsymbol{\sigma}
\vert m_{0},-m_{B},m_{B}\right)$ where $\omega\left( \boldsymbol{\sigma} \rightarrow \boldsymbol{\sigma'}
\right)=\rm{min}\{1,\exp(-\beta \Delta H)\}$ is the Metropolis success rate
in coarse grain time variable $\tau$ and $P\left(\boldsymbol{\sigma}
\vert m_{0},-m_{B},m_{B}\right)$ is the steady state probability of
configuration $\boldsymbol{\sigma}$ corresponding to
subsystem magnetization $m_{0}$, $m_{u}=-m_{B}$ and
$m_{\ell}=m_{B}$. Summing over all such exchanges one obtains
\begin{equation}
p\left(
m_{0}\right)=\sum_{\boldsymbol{\sigma},\boldsymbol{\sigma^{\prime}}}\omega\left(
\boldsymbol{\sigma}\rightarrow\boldsymbol{\sigma^{\prime}}\right)P\left(\boldsymbol{\sigma}\vert
m_{0},-m_{B},m_{B}\right),
\label{eq:general p}
\end{equation}
where the sum is over configurations $\boldsymbol{\sigma^{\prime}}$ whose $m_{0}$
is higher than that of $\boldsymbol{\sigma}$ by $2/L$.
The magnetization decreasing rate $q(m_{0})$ is readily obtained by
noting that due to the invariance of the dynamics to space-time
inversion, $\sigma_{\mathbf{r}}\rightarrow -\sigma_{-\mathbf{r}}$, one
has $q(m_{0})=p\left(-m_{0}\right)$.

In the slow exchange limit $\gamma\ll L^{-3}$, the probability $P\left(\boldsymbol{\sigma}\vert
m_{0},-m_{B},m_{B}\right)$ can be expressed in terms of probability of
the subsystem configurations as
\begin{eqnarray}
P(\boldsymbol{\sigma} \vert
m_0,-m_{B},m_{B})\simeq&P(\boldsymbol{\sigma}_{0}\vert m_{0})
P\left(\boldsymbol{\sigma}_{u}\vert-m_{B},m_{0}\right)\nonumber\\
&P\left(\boldsymbol{\sigma}_{\ell}\vert m_{B},m_{0}\right),
\label{eq:adiabatic}
\end{eqnarray}
where $\boldsymbol{\sigma}_{0}$, $\boldsymbol{\sigma}_{u}$ and
$\boldsymbol{\sigma}_{\ell}$ are the microscopic spin configurations of the three subsystems
corresponding to the configuration
$\boldsymbol{\sigma}$. Here, $P(\boldsymbol{\sigma}_{0}\vert m_{0})$ is the steady state
distribution of the driven line with fixed magnetization $m_{0}$,
which is the same as the steady state of a TASEP, and
$P\left(\boldsymbol{\sigma}_{u}\vert-m_{B},m_{0}\right)$ and
$P\left(\boldsymbol{\sigma}_{\ell}\vert m_{B},m_{0}\right)$ are the equilibrium
distribution of the other two subsystems.
\begin{figure}
\begin{center}
\includegraphics[width=5.0cm]{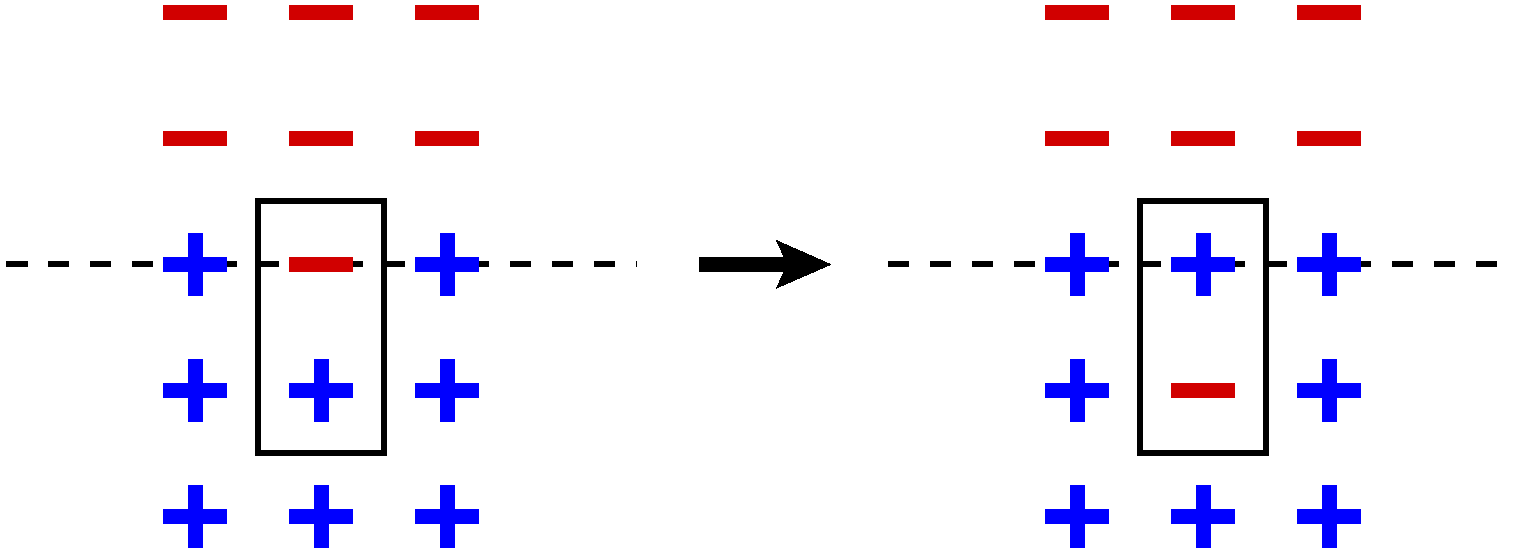}
\caption{(color online) Spin exchange process which yields the leading
order contribution to $p(m_0)$. The driven ring is denoted by a dotted
line, and the pair of
spins exchanged are indicated. \label{fig:Two-examples-of}}
\end{center}
\end{figure}

In general, calculating all the terms in Eq. \eqref{eq:general p} is
not straightforward. However,
the calculation becomes feasible in the low $T$ limit where these rates may
be expanded in powers of $\exp\left(-\beta J\right)$.

In order to keep track of the terms in this expansion it is convenient to
generalize the model by considering
an interaction strength between the driven ring and its neighboring
ones as $J_{1} \le J$. It is easy to see that the leading contribution to
$p(m_{0})$ in Eq. \eqref{eq:general p} results from the exchange process
shown in Fig. \ref{fig:Two-examples-of}, where both subsystems $u$ and $\ell$ are in
their respective ground state, $m_{\ell}=-m_{u}=1$. For this process $\omega\left(
\boldsymbol{\sigma}\rightarrow\boldsymbol{\sigma^{\prime}}\right)=\exp(-2\beta(J+J_{1}))$
and $P\left(\boldsymbol{\sigma}\vert m_{0},-m_{B},m_{B}\right)=\left[\left(1+m_{0}\right)^{2}\left(1-m_{0}\right)/8\right]\left[1-\mathcal{O}\left(e^{-6\beta
J}\right)\right]$. Higher order contributions can be determined
similarly from other exchange events. Computing $p(m_{0})$ up to
$\mathcal{O}(\exp(-6\beta J))$ yields, for $-1 \le m_{0}\le 1$,
\begin{multline}
p\left(m_{0}\right)=\frac{1}{8}\left[\left(1+m_{0}\right)^{2}\left(1-m_{0}\right)e^{-2\beta
J_{1}}\right ]e^{-2\beta J}\\
+\frac{1}{8}\Bigl[\left(1+m_{0}\right)^{2}\left(1-m_{0}\right)\left(2
e^{-2\beta J_{1}m_{0}}+e^{2\beta J_{1}m_{0}}\right)\\+2\left(1+m_{0}\right)\left(1-m_{0}\right)^{2}\left(e^{-2\beta J_{1}}+e^{2\beta J_{1}m_{0}}\right)\\
+\left(1-m_{0}\right)^{3}e^{2J_{1}m_{0}}\Bigl]e^{-6\beta J}
+\mathcal{O}\left(e^{-8\beta J}\right).
\label{eq:rate}
\end{multline}

The LDF $\phi(m_{0})$ calculated using the rate in Eq. \eqref{eq:rate}
is plotted in Fig. \ref{fig:The-potential-function} for $\beta J_{1}=\beta J=3/4$ ($T\simeq0.6T_{c}$).
This function has two minima which correspond to the two thermodynamic phases with non-zero
$m_{0}$. The average time $\tau_{s}$ for the magnetization to jump
from one minimum to the other is proportional to the exponential of the
barrier height between them. In terms of Monte Carlo
steps this switching time $t_{s}\sim \gamma^{-1}\tau_{s}\sim \gamma^{-1}\exp(\epsilon L)$
where $\epsilon$ is the barrier height. For the parameters of Fig.
\ref{fig:The-potential-function} one has $\epsilon=0.18$ which is of the same order as that
obtained numerically in Fig. \ref{fig:Exponential-growth-of}. For a better
comparison higher order terms in the low temperature expansion are required. The asymmetry in the
fluctuations of the interface and the magnetization profile in Fig.
\ref{fig:The-average-magnetization} is a consequence of the non-zero values of
$m_{0}$.

The analysis presented in this Letter demonstrates that a local drive can induce a
phase transition which involves spontaneous symmetry breaking
of an interface separating two coexisting phases. It would be interesting to
consider other boundary conditions which would allow the interface to detach from the
driven ring. This would correspond, for example, to studying the model with periodic boundary conditions
in both the $x$ and $y$ directions. In this case the model exhibits
two interfaces, and preliminary studies have shown that either one of
them is attracted by the driven ring resulting in a
macroscopic symmetry breaking, in addition to that of the interface \cite{ZVI}. This will be addressed
in a future publication.

\begin{acknowledgments}
We thank A. Bar, O. Cohen, M.R. Evans, O. Hirschberg, S. N. Majumdar, A.
Macio\l ek and S. Prolhac for helpful discussions.
The support of Israel Science Foundation (ISF) is gratefully acknowledged.
\end{acknowledgments}
\bibliography{reference}
\bibliographystyle{apsrev4-1}
\end{document}